\documentclass[manuscript,screen]{acmart}

\AtBeginDocument{%
  \providecommand\BibTeX{{%
    \normalfont B\kern-0.5em{\scshape i\kern-0.25em b}\kern-0.8em\TeX}}}

\setcopyright{acmcopyright}
\copyrightyear{2020}
\acmYear{2020}
\acmDOI{XX.XXXX/XXXXXX.XXXXXXX}

\acmJournal{TOCHI}
\acmVolume{1}
\acmNumber{1}
\acmArticle{1}
\acmMonth{1}

\usepackage{graphicx}
\usepackage{booktabs}
\usepackage{layouts}  
\usepackage{subcaption}
\usepackage{hyperref}
\usepackage{tabularx}
\usepackage{array}

\usepackage{booktabs}
\usepackage{multirow}
\usepackage{longtable}
\usepackage{caption}
\usepackage[utf8]{inputenc}
\usepackage[inline]{enumitem}

\definecolor{ForestGreen}{rgb}{0.11, 0.46, 0.11}
\newcolumntype{C}[1]{>{\centering\arraybackslash}p{#1}}

\newcommand{\quotes}[1]{``#1''}
\begin{document}

\title{Comparing Generic and Community-Situated Crowdsourcing for Data Validation in the Context of Recovery from Substance Use Disorders}

\author{Sabirat Rubya}
\email{rubya001@umn.edu}
\orcid{0000-0002-8932-1489}
\affiliation{%
  \institution{University of Minnesota}
  \streetaddress{...}
  \city{..}
  \state{..}
  \postcode{..}
}

\author{Joseph Numainville}
\email{numai004@umn.edu}
\affiliation{%
  \institution{University of Minnesota}
  \streetaddress{..}
  \city{..}
  \state{..}
  \postcode{..}
}

\author{Svetlana Yarosh}
\email{lana@umn.edu}
\affiliation{%
  \institution{University of Minnesota}
  \streetaddress{..}
  \city{..}
  \state{..}
  \postcode{..}
}

\renewcommand{\shortauthors}{Author1LastName et al.}

\begin{abstract}
  Targeting the right group of workers for crowdsourcing often achieves better quality results. One unique example of targeted crowdsourcing is seeking community-situated workers whose familiarity with the background and the norms of a particular group can help produce better outcome or accuracy. These community-situated crowd workers can be recruited in different ways from generic online crowdsourcing platforms or from online recovery communities. We evaluate three different approaches to recruit generic and community-situated crowd in terms of the time and the cost of recruitment, and the accuracy of task completion. We consider the context of Alcoholics Anonymous (AA), the largest peer support group for recovering alcoholics, and the task of identifying and validating AA meeting information. We discuss the benefits and trade-offs of recruiting paid vs. unpaid community-situated workers and provide implications for future research in the recovery context and relevant domains of HCI, and for design of crowdsourcing ICT systems.
\end{abstract}

\ccsdesc[500]{Human-centered computing~Human computer interaction (HCI)}

\keywords{Crowdsourcing, Alcoholics Anonymous, community-situated crowd}

\maketitle


\section{Introduction} 
In recent years crowdsourcing systems have widely been used in both industry and academia for collecting human-labeled data. Generic crowdsourcing platforms like Amazon Mechanical Turk attract a large number of workers. However, since enrollment to such platforms does not require any particular skill set from the workers, ensuring the quality of the crowdsourced data is often a challenge. Attracting workers who are likely to have the skills needed for the target task may be useful to overcome this problem. Previous research has shown effectiveness of crowdsourcing with the \quotes{right workers} (e.g., who are present at a particular location or time, or have familiarity with or expertise in performing particular tasks) \cite{crisisdisaster, Jonathan2018Stella:,Vries2017Experts, almostexpert}. Building and expanding on this idea, we aim to understand if the unique perspectives and contextual knowledge of members of a community relevant to the target task can help achieve better quality results in crowdsourcing.

There are different techniques to recruit and filter potential community members. They can be recruited from different platforms or online communities, and with or without payment. In general, unpaid crowdsourcing leads to a potentially unpredictable workforce and indeterminable task completion time due to the lack of financial incentive\cite{unpaidhigh}. Paid tasks, on the other hand, can attract crowd workers who may falsely report their skill, confidence, or expertise required to complete the tasks \cite{zhu10, bruhlmann_quality_2020}. In other words, there are performance trade-offs in crowdsourcing in terms of task completion time, cost, and accuracy, while applying different crowd-work techniques to recruit generic vs. community members from the same paid platform, and community-situated crowd through paid vs. unpaid platforms. We contribute an empirical understanding of these trade-offs which bear significance in a variety of contexts where applying specific recruitment techniques or targeting specific platforms to access potential community-situated crowd workers can provide better accuracy of completing particular tasks. For instance, members from an online community of autism caregivers can provide more concise and useful information and advice to people with autism to cope with everyday challenges, due to having more experience than the non-members. Similarly, citizen science projects can benefit from crowdsourcing data validation from online platforms consisting of people who have domain expertise. In this paper, we analyze these performance trade-offs by answering the following research questions:

\begin{itemize}
    \item {\emph{\textbf{RQ1:}} How do self-reported community-situated crowd workers perform compared to generic crowd workers in terms of task completion time, accuracy, and cost? }
    \item {\emph{\textbf{RQ2:}} What are the benefits and trade-offs of recruiting potential community-situated crowd workers with or without payment and screening techniques? }
\end{itemize}

We analyze these trade-offs in the context of data validation tasks to create a peer-support meeting list for Alcoholics Anonymous (AA). AA has over 1.5 million members and hosts over 100 thousand weekly groups worldwide \cite{Ferri2006Alcoholics}. AA focuses on providing in-person meetings where people share and provide support to each other. People who are recovering from alcoholism rely on forming and engaging in these mutual-help groups in order to stay sober. While AA meeting attendance is particularly important for people who are newly sober, newcomers often have trouble finding reliable information about meeting locations and times due to a lack of a global up-to-date meeting list, a problem that arises due to a preference for regional autonomy in AA’s organizational structure \cite{Yarosh2013Shifting, Rubya2017Facilitating}.

Previously, researchers attempted to make the meeting information available and up-to-date in a \quotes{global
meeting list} through detection of regional AA websites containing meetings and the extraction of day, time, and address of meetings from those sites. They adopted a human-aided information retrieval approach for this purpose where automated machine learning and pattern detection approaches extract information about possible meetings listed on different AA websites. Then, these meeting webpages and meeting information were validated through paid human workers on an online crowdsourcing platform \cite{HAIR}. We refer to this crowdsourcing technique, where a wide variety of \quotes{crowds} (without a particular set of assumed skills) can be recruited online or offline to complete tasks, as \quotes{generic crowdsourcing.} However, a major source of error in this technique came from the workers providing poor quality answers and failing to identify webpages with meetings listed on them and content that provided information about other AA events rather than actual AA meetings.

The regional AA websites are in different formats and structures and AA has particular norms and traditions rooted in the program. Members of this community, who attend the AA meetings in person, may have a unique perspective on these norms that is different from non-members. In addition, many of them are more familiar with the organization and content of the meeting websites. However, approximately 1.3 million US residents (0.41\% of US population) are AA members, and a random pool of workers recruited from a generic online platform will possibly constitute very few or no community members. Therefore, one reason of the poor performance of the generic crowd workers may be due to the lack of their contextual knowledge. In addition to the familiarity with the program norms and the meeting websites, members of AA and similar peer-support group are motivated to perform service work to help other newcomers in the community \cite{Rubya2017Video-Mediated}. We aim to understand if the unique perspectives and contextual knowledge of the AA members and their intrinsic motivation to help their peers can achieve better quality results to crowdsource accurate meeting information. In this regard, we define \quotes{the technique of seeking potential workers whose membership in particular communities of practice provides them with a unique perspective on relevant background and norms of that community} as \quotes{community-situated crowdsourcing}.

To compare generic and community-situated crowdsourcing, we recruit workers in three different techniques: 1) crowd workers from a paid online crowdsourcing platform (Amazon Mechanical Turk) without any particular screening techniques (paid unfiltered generic), 2) crowd workers from the same paid platform but only allow workers who self-report themselves to be 12-step fellowship members and are able to answer a few basic screening questions (paid filtered self-reported community-situated), and 3) members from an online community for recovery from substance abuse (unpaid filtered special community members), and evaluate these techniques of filtering crowdworkers in terms of time, accuracy and cost. We found that community-situated workers recruited from an online community can achieve better accuracy in AA meeting information identification and validation than paid generic crowd workers from Amazon Mechanical Turk, though it may take substantially longer to recruit community-situated crowd workers from such online community, particularly if they are recruited as unpaid volunteers. Additionally, we found evidence in our data pointing out that the community-situated workers' expertise and familiarity with the context may have helped them in achieving better accuracy. From these findings, we provide implications for effectively filtering and utilizing community-situated crowd workers in this and other domains, as well as discuss the implications in the broader context of other crowdsourcing ICT systems.

The contributions of this paper include:
\begin{enumerate}[topsep=3mm]
    \item{an empirical comparison of  crowdsourcing from unfiltered generic, self-reported filtered community-situated, and self-reported filtered special community members in terms of time of recruitment, cost, and accuracy, and understanding the tradeoffs involved in utilizing these different forms of crowd work,}
    \item{adding to our understanding of the performance benefits of using community-situated crowdsourcing for specific cases, in contrast to generic crowdsourcing methods. Crowdsourcing studies have recommended community-situated crowdsourcing as more optimal than generic crowdsourcing due to perceived benefits by the community group \cite{quizz, Wauck2017From}, but not a lot of work has been done to show these performance benefits in practice. This work quantitatively establishes the benefits of community-situated crowdsourcing with an example context of recovery.}
\end{enumerate}

We begin by situating our work in the context of previous research on situated and community-based crowdsourcing and on the comparison of different types of crowdsourcing. Next, we describe the methods and the findings of our analysis of time, cost, and accuracy for crowdsourcing the tasks related to AA meeting validation with generic unfiltered vs. self-reported filtered workers from a generic online crowdsourcing platform, and with self-reported online community members. We conclude with a discussion connecting this work to the broader domain of HCI and CSCW by providing implications for research on crowdsourcing and for design in other ICT systems.

\section{Related Work} 

\subsection{Situated Crowdsourcing}

 In the case of generic crowdsourcing, crowd feedback is solicited from crowds driven by financial motivations \cite{Yen2016Social}. However, crowd workers can be motivated by intrinsic aspects other than the monetary remuneration, such as enjoyment, or community based motivation \cite{Hossain2012Users', Kaufmann2011More}.

While online crowdsourcing markets make it convenient to pay for workers willing to solve a range of different tasks, they suffer from limitations such as not attracting enough workers with desired background or skills \cite{Li2014Wisdom, Alonso2011Crowdsourcing, Erickson2012Hanging}. For example, it can be a challenge to recruit workers for a task that requires workers who speak a specific language or who live in a certain city \cite{Jonathan2018Stella:, Altruistic} . Situated crowdsourcing can help fill in the gaps in these scenarios where the crowd needs to be associated with some context. Situated crowdsourcing consists of embedding input mechanisms (e.g., public displays, tablets) into a physical space and leveraging users’ serendipitous availability \cite{gonsalvessituatedcrowd} or idle time (\quotes{cognitive surplus}) \cite{Shirky2010Cognitive}. It allows for a geo-fenced and more contextually controlled crowdsourcing environment, thus enabling targeting certain individuals, leveraging people's local knowledge or cognitive states, or simply reaching an untapped source of potential workers \cite{Goncalves2013Crowdsourcing, Goncalves2015Bazaar:, Goncalves2014Game, Hosio2014Situated}. Researchers have discussed benefits of targeting geographically or temporally situated crowds over generic crowds in scenarios like providing emergency services in disasters \cite{Ludwig2017Situated}, geotagging photos \cite{Jonathan2018Stella:}, etc. An experiment by Ipeirotis et al. automatically identified \quotes{situated crowds} with desired competence to complete a task and demonstrated that the cost of hiring workers through their platform is less than that of hiring workers through paid crowdsourcing platforms \cite{quizz}.  However, these previous works do not focus on situated crowdsourcing where competence of the workers for the target task has resulted from being members of a particular community, nor do they consider recruiting the targeted workers from different platforms. We extend the idea of situated crowdsourcing to capture the context of selecting the \quotes{right crowd} who are better suited for a task (e.g., with a particular skill or quality) due to their familiarity with the context, are more reliable, and can provide better quality results \cite{Curmi2015Crowdsourcing, Li2014Wisdom}.

Most prior work focuses on temporally-situated or locally-situated participants, who are at a time or place to best be able to do the task. Building on these ideas, we refer to the process of recruiting potential crowd-workers who are members of particular communities having better context-knowledge and who may be motivated to produce better results without monetary incentives as \quotes{community-situated} crowdsourcing. Some examples include detecting and reporting predatory publishers through crowdsourcing from the community of authors and researchers \cite{publisherdetector}, or getting feedback for course projects on a particular course from a community of freelance experts on that topic \cite{almostexpert}. In the context of this paper, members of 12-step communities like AA are familiar with the format of recovery meetings and governing structure meetings. A major source of error in previous work with crowdsourcing AA meeting validation \cite{HAIR} came from workers being unable to distinguish area or district meetings from open weekly AA meetings. However, we can assume that actual members of this community would not make the same mistakes, thus being the \quotes{right crowd} with the required skill (familiarity with the context) for the tasks of 12-step meeting identification and validation.

The primary contribution of our work is empirically comparing two alternative approaches for community-situated crowdsourcing. The secondary contribution is quantitatively establishing the benefits of community-situated crowdsourcing in the context of recovery.

\subsection{Community-Situated Crowdsourcing: Generic Platforms versus Targeted Online Communities}

There may be multiple reasonable approaches for recruiting a community-situated crowd. For example, when using a generic platform like Amazon Mechanical Turk, membership in a particular community can be self-reported and can serve as a qualification for completing the task, although this process does not ensure that all the recruited participants are community members. Alternatively, one may be able to solicit members of a desired group directly through online spaces dedicated to those communities (e.g., Facebook recovery groups, InTheRooms.com). Prior work has incorporated elements of both approaches, but without systematically separating and comparing the two. For example, friendsourcing leverages one's social network to gather information that might be unavailable or less trustworthy if obtained from other sources \cite{Bateman2017Comparing, Bernstein2008Personalization, Martins2014Friendsourcing, Rzeszotarski2014Estimating}. Researchers have attempted to build systems that, for example, use friendsourcing for social tagging of images and videos \cite{Bernstein2008Personalization}, to seek out personalized recommendations, opinions, or factual information \cite{Bernstein2008Personalization, Morris2014Remote}, etc., to help blind communities get answers to questions about the world around them captured by cameras \cite{Brady2013Investigating}, or to provide cognitive aids to people with dementia \cite{Martins2014Friendsourcing}.

Friendsourcing removes the financial cost of the service and improves the quality and trustworthiness of the answers received \cite{Gonzalez2013Recommendations, Morris2010Comparison}. Several studies applying friendsourcing in health contexts (e.g., dementia, Alzheimer’s disease, etc.) suggest that friendsourcing can encourage engagement and provide more emotional and informational support compared to crowdsourcing \cite{Bateman2017Comparing}. Friendsourcing is a specific example of leveraging an existing known community and soliciting members through existing or novel online channels to complete desired tasks.

As another example, altruistic crowdsourcing refers to cases where unpaid tasks are carried out by a large number of volunteer contributors \cite{Altruistic}. This form of crowdsourcing often utilizes members of the same community to complete collective tasks or getting better quality and more trustworthy information \cite{Altruistic, Goncalves2013Crowdsourcing}. However, some situations require altruistic contribution from out-group crowd workers, for example, in providing directions to people with visual impairments \cite{Brady2015Getting}, or in generating valuable daily life advice for people with autism \cite{Hong2015In-group}. Altruistic crowdsourcing may leverage existing social networks or may serve as a filter when seeing contributions from the larger community (in the sense that those willing to complete unpaid tasks in a particular context are likely to have a personal connection or interests in that context). Both friendsourcing and altruistic crowdsourcing approaches rely heavily on social motivators such as social reciprocity, the practice of returning positive or negative actions in kind, reinforcing social bonds, the opportunity of showing off expertise, etc. \cite{Bernstein2008Personalization, Curmi2015Crowdsourcing, Brady2013Investigating}.

Systems that utilize members of the same community or known channels as the crowd often leverage their intrinsic interest in a particular domain and their sense of belonging to the community \cite{Kaufmann0Worker, Schenk2011Towards}. This form of crowdsourcing has proven to receive better quality feedback for class projects from classroom peers \cite{Wauck2017From, Alfaro2014CrowdGrader:}. Moreover, researchers have discussed the efficacy of crowdsourcing peer-based altruistic support in critical contexts, such as to reduce depression and to promote engagement \cite{Morris2015Efficacy}, for generating behavior change plans \cite{Agapie2016PlanSourcing:, Vries2017Experts}, to exchange health information \cite{Rubenstein2013Crowdsourcing}, etc. Although friendsourced answers often contain personal or contextual information that improves their quality, in some cases people do not consider social network as an appropriate venue for asking questions due to high perceived social costs, limiting the potential benefit of friendsourcing \cite{Rzeszotarski2014Estimating, Bateman2017Comparing, Brady2013Investigating}.

This body of work provides compelling examples of benefits of community-situated crowdsourcing but is largely opportunistic about how such a situated crowd is recruited. Obviously, friendsourcing could not be reasonably accomplished through filtering workers on generic platforms. However, in most other cases, both qualifying members on existing platforms and targeting specific online communities may be reasonable approaches and may have different costs and benefits. Our primary contribution is providing an empirical comparison of accuracies, costs, and time trade-off in these two approaches to community-situated crowdsourcing, by comparing tasks completed by workers on MTurk who self-identify as members of 12-step programs and those done by members of the targeted online recovery community InTheRooms. Based on our findings, we provide recommendations for community-situated crowdsourcing in other contexts.

\subsection{Comparison of Different Types of Workers}

Both CHI and CSCW communities have studied the comparison among different types of crowdsourcing based on magnitudes of financial incentives \cite{targetedqa, mao2013a, yin2013a}, worker motivations \cite{rogstadius2011a, socialincentive}, worker expertise \cite{citizenscience,citizenscience2, Vries2017Experts}, etc.

Many initial crowdsourcing studies focused on finding out how the magnitude of financial incentives affects the work produced. While some of the studies suggested that worker quantity may increase with higher incentives for the same task but the quality of the results do not improve \cite{mao2013a, mason2009, tightbudget}, others pointed out that the amount of monetary incentives can produce better quality work if it is performance-contingent (e.g., rewards or penalties) \cite{yin2013a}. On the other hand, unpaid crowd workers provide a workforce without labor cost and can work as an economical alternative for individuals and organizations who are concerned about the budget \cite{borromeo2016investigation, unpaidhigh}. Researchers pointed out that volunteers often provide as reliable and high quality answers as paid workers, though the turnaround time may be higher for unpaid workers and they are more likely to not complete the tasks when compared to their paid counterparts \cite{unpaidhigh, mao2013a, redi2014, pledgework}, making the use of crowdsourcing through volunteers questionable for time-sensitive tasks.

Since non-expert crowd workers are often more cost-effective than expert ones, another body of research studied the performances of crowdsourcing by experts and non-experts. They suggested that non-expert work quality may be comparable to the experts \cite{mellebick,gameelements} in many contexts, including tasks as difficult as identifying a particular type of bio signal (sleep-spindle) from raw data \cite{warby2014a}. Moreover, research on leveraging these types of workers in citizen science, web security, etc. suggested that their complementary roles and different potentials in different tasks should be better capitalized by community-based systems across different domains \cite{citizenscience,websecurity, communitycontribute}.

While some of these research work discuss the benefits of recruiting expert and reliable volunteers from specific communities, they do not explicitly compare performances of these participants recruited with different approaches or from different platforms. In peer-support communities for critical health conditions, there may be urgency regarding time and accuracy of the produced results for crowdsourcing tasks \cite{Hong2015In-group, wazny_applications_nodate}. We explicitly compare community-situated paid and unpaid crowdsourcing in the high-impact context of recovery and discuss trade-offs that can be applied to other similar contexts.

\section{Methods}
We conducted an experiment to understand the trade-offs of generic vs. community-situated crowdsourcing. For generic crowdsourcing, we recruited workers from Amazon Mechanical Turk (MTurk) and for community-situated crowdsourcing we solicited potential community members from both a filtered subset of MTurk and an online community for people in recovery. We performed statistical comparisons to understand the differences in performances of the workers recruited with these three different techniques.

\subsection{AA Meeting Dataset}
The steps of developing a comprehensive AA meeting list include identifying different regional webpages on different domains that contain one or multiple AA meetings, and locating all the meetings and their corresponding information (e.g., day, time, and address) on those meeting pages. To create such a list, researchers previously paired information retrieval with human computation \cite{HAIR} and collected \begin {enumerate*} [label=\itshape\alph*\upshape)] \item{ ground truth data labels for $964$ webpages, each with a label of 1 or 0 indicating whether it contains a list of meetings or not}, and \item{location of $1892$ meetings from a subset of the meeting pages. This subset was selected from 18 regional websites from three different states in the USA.}\end{enumerate*} We got access to this data from prior work and saved them in a MySQL database on a secure server obtained from our institution.

\subsection{Participants}\label{sec:participants}
We recruited three types of participants: 1) crowd workers from MTurk without any particular screening techniques (generic MTurk), 2) crowd workers from MTurk but only allow workers who self-report themselves to be 12-step fellowship members and are able to answer our screening questions (community-situated MTurk), and 3) volunteers from an online community for recovery from substance abuse (community-situated InTheRooms). We collected data for the first type from researchers of previous work. Since we already had data for one of the three populations (generic MTurk) and we wanted to minimize the cost of recruiting workers, we utilized the collected data to perform a power analysis. We found  that for our statistical comparisons the desired sample sizes (the minimum number of participants) for the other two populations (community-situated MTurk and community-situated InTheRooms) should be 400 to achieve a power of 0.8 at the 95\% confidence level.

\subsubsection{Generic Crowd Workers from MTurk}

Amazon Mechanical Turk or MTurk\footnote{\href{https://mturk.com}{https://mturk.com}} is a popular online crowdsourcing marketplace where \quotes{requesters} can hire remotely located \quotes{crowd workers} (MTurkers) to perform discrete on-demand tasks known as Human Intelligence Tasks (HITs). The \quotes{generic crowd} sample in the previous work consisted of a total of 1060 individuals recruited from MTurk \cite{HAIR}. They created a HIT with the background and description of their study along with a link of the website to view and agree to an online informed consent and to complete the tasks. After they completed all three tasks, they were provided with a unique survey code which they had to copy and paste on the MTurk platform so their responses could be tracked and associated with corresponding worker IDs to approve or disapprove their HITs. Participants were recruited during the month of May 2018.

\subsubsection{Filtered Community-situated Crowd Workers from MTurk}
For recruiting and filtering MTurk workers who are practicing their recovery through one or more 12-step fellowships, we designed a screening survey and explicitly mentioned that only members of 12-step programs should attempt the questions. The screening survey consisted of three basic questions about recovery practices in 12-step fellowships (see below). 660 workers attempted the survey, and 480 of them answered all three questions correctly. We accepted all the HITs for the screening survey, but only these 480 workers were marked as \quotes{qualified} workers. The actual HIT with the link to the tasks (described in Section \ref{sec:tasks}) was made visible only to the qualified workers. Participants were recruited during the month of December 2019.

\textit{Screening Survey Questions:}
The screening survey included the following three questions about recovery in 12-step fellowships (Fig 2 in supplementary document):
\begin{itemize}
    \item{What is your primary 12-step fellowship?}
    \item{How many meetings do you attend weekly?}
    \item{Fill in the blank: The 12th tradition says, \quotes{\_\_\_\_\_\_\_\_\_ is the spiritual foundation of all our Traditions, ever reminding us to place principles before personalities.}}
\end{itemize}

\textit{Ensuring Quality of the Workers:}
Previous work has pointed out that paid crowd workers recruited from online platforms often produce low quality results \cite{maliciousworker} to maximize their earnings by completing tasks quickly. In order to assure high quality recruitment of participants, we used several strategies. First, the recruitment was limited to MTurkers whose average HIT acceptance rate was 90\% or higher. The second strategy was to embed a custom script on the HIT page that limits the number of times that a single worker may work on this study. By implementing this code, MTurkers could take part in this survey only once, reducing the effect of noise in the results from worker experience over repeated tasks. Lastly, all the response patterns were reviewed on a daily basis and all speeders and straight-liners were removed every day. For instance, if a respondent completed our survey too quickly (i.e., speeders who took less than five minutes in total starting from reading the consent form to finishing the last task) without carefully reading the question items, the responses from that respondent were excluded from the analysis. Also, if a respondent rushed through the tasks by clicking on the same response to multiple questions (i.e., straight-liners for all ten different answers for a particular task), those cases were also filtered out from data analysis. As a result of such data cleaning procedures, different participant answers were removed for different tasks, which resulted in an unequal number of participants for three different tasks. For community-situated MTurkers, the sample sizes for page validation, meeting validation, and meeting identification were 423, 406, and 435 respectively.

It is an important point to note again that the term \quotes{community-situated crowdsourcing} is a shorthand for the process of seeking workers who are members of particular groups or communities related to the context of the research questions. All the community-situated workers in our study have \textit{self-reported} themselves to be members of 12-step fellowships. We cannot, however, make a certain claim about all of them actually being in recovery. MTurkers can lie about being AA members and bypass the screening questions, and we discuss later about how that can reflect in their performances. We kept the screening survey questions simple so it does not take too long to recruit target number of \quotes{self-reported filtered community-situated crowd workers} from MTurk.

\subsubsection{Community-situated Crowd Workers from ITR}
\label{sec:itrworkers}
InTheRooms.com\footnote{\href{https://intherooms.com}{https://intherooms.com}}(ITR) is the largest online community for recovering addicts, and their friends and families, hosting over 500 thousand members. The website is a free public social networking site focused on connecting people in recovery and others affected by substance use disorders with peer support. It hosts over 100 weekly online video-meetings and provides social features for members to connect with and provide support to each other (e.g., profiles, wall posts, and comments).

We worked with the ITR website founders and owners to reach out directly to ITR members to participate in the study. A link to the survey was distributed as a paid banner advertisement on the ITR homepage for two months (October 2019-November 2019), as well as advertised in their weekly ITR newsletter with a message from the researchers explaining the purpose of the study (Fig. 1 of supplementary document). Similar to the workers from MTurk, all ITR members who responded to the advertisement viewed and agreed to an online informed consent.

Again, the community-situated crowd workers from ITR were recruited based on the assumption that the online community members are in fact 12-step fellowship members, which may not be true for all members of ITR, since previous work has established that online communities may have spammers or non-members \cite{spammer1, spammer2}. However, based on a few assumptions from our research experience with this community for past couple years, we think that unlike MTurk, ITR participants are more likely to be actual 12-step fellows, and hence we did not ask the filtering questions to the ITR members. First, ITR is primarily an online peer support network for 12-step fellowship members and their friends and families. A spammer in this community does not have much to gain from frequent interactions with others. Second, even if we assume that there is a considerable number of non-12 step fellowship members in the community, the probability of them responding to a study without any monetary reward should be very low. Finally, since we were not providing any reward to the ITR members for participating in the survey, providing additional questions for screening might have reduced the turnover of the responses, increasing the time and the cost of recruitment.

\subsubsection{Ethical Considerations}
Recovery from substance use is a personal and private undertaking for most people and we took steps to consider the ethical implications of our work and to protect the rights of the community-situated workers who are members of different 12-step programs. All research activities on this project were reviewed and approved by our university IRB as an investigation where potential benefits of the scientific work outweighed the risks to the participants. The approval covered two phases. First, all users who completed the tasks viewed and responded to an online informed consent form, but documentation of informed consent was waived in order to preserve participants' anonymity. Second, we made sure that the screening survey and the tasks do not ask the participants for any personally identifiable information.

\begin{figure}
    \centering
   \includegraphics[width=.85\textwidth]{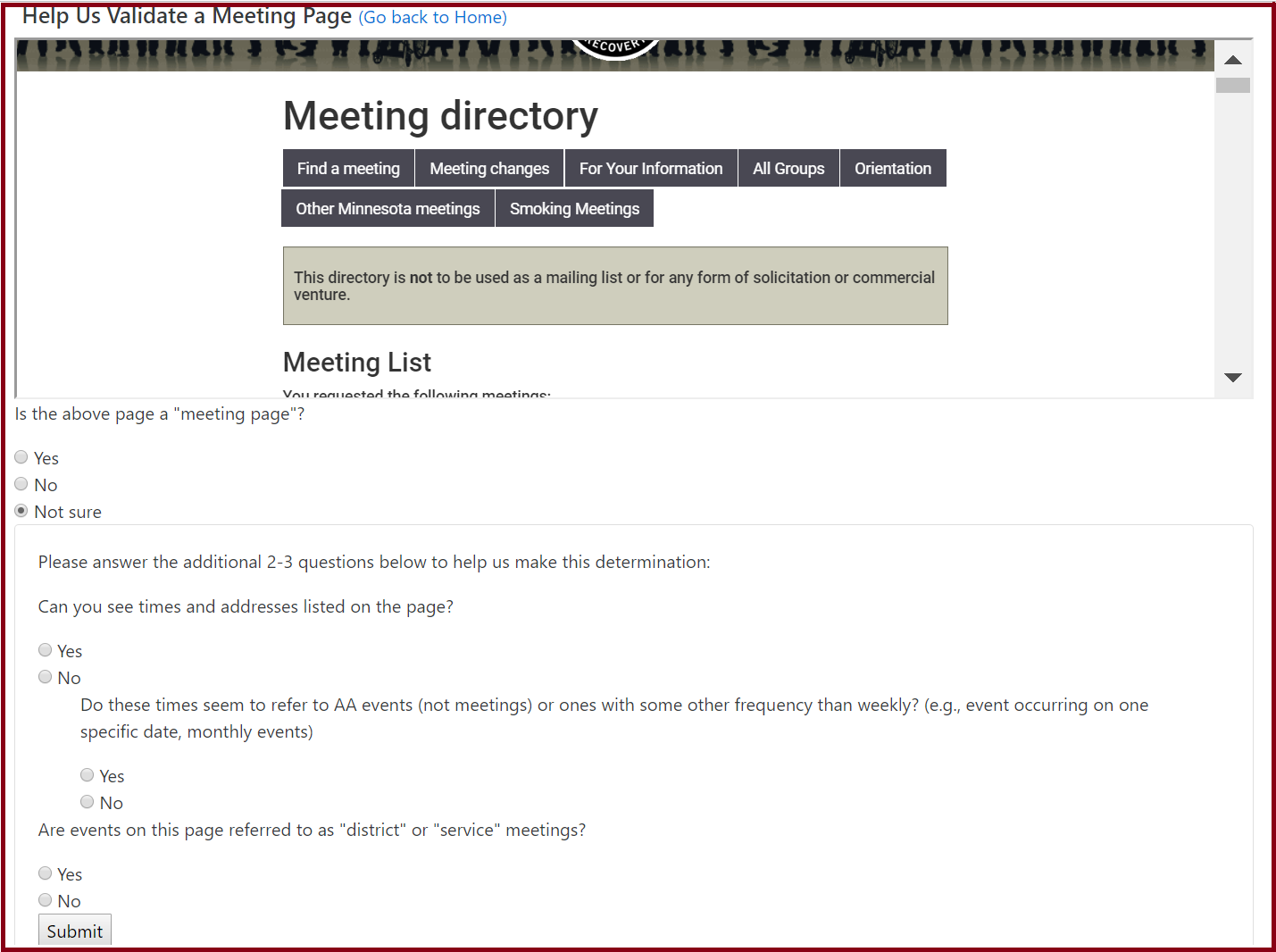}
    \caption*{(a)}
   \includegraphics[width=.85\textwidth]{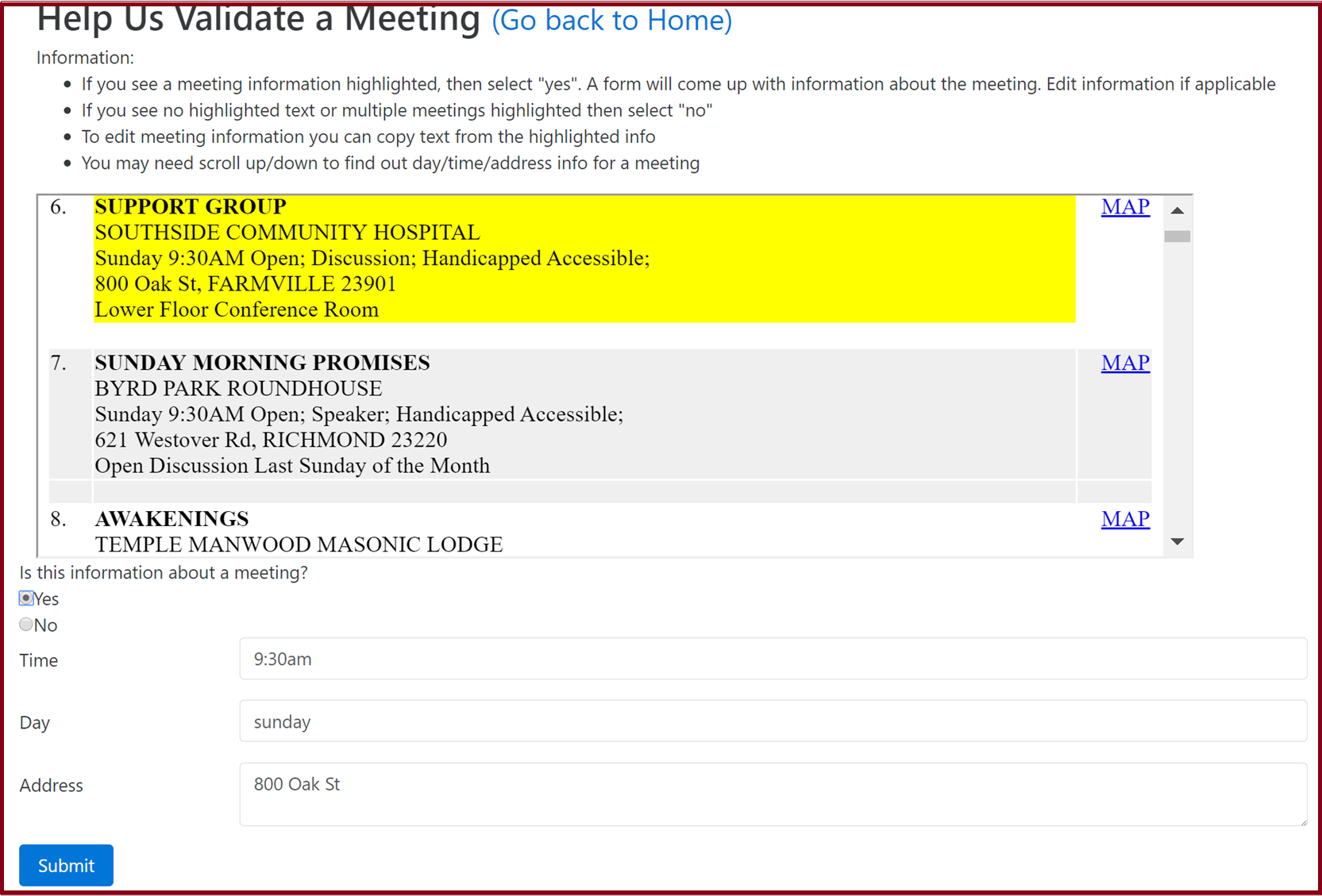}
    \caption*{(b)}
\end{figure}
\begin{figure}
    \centering
   \includegraphics[width=.85\textwidth]{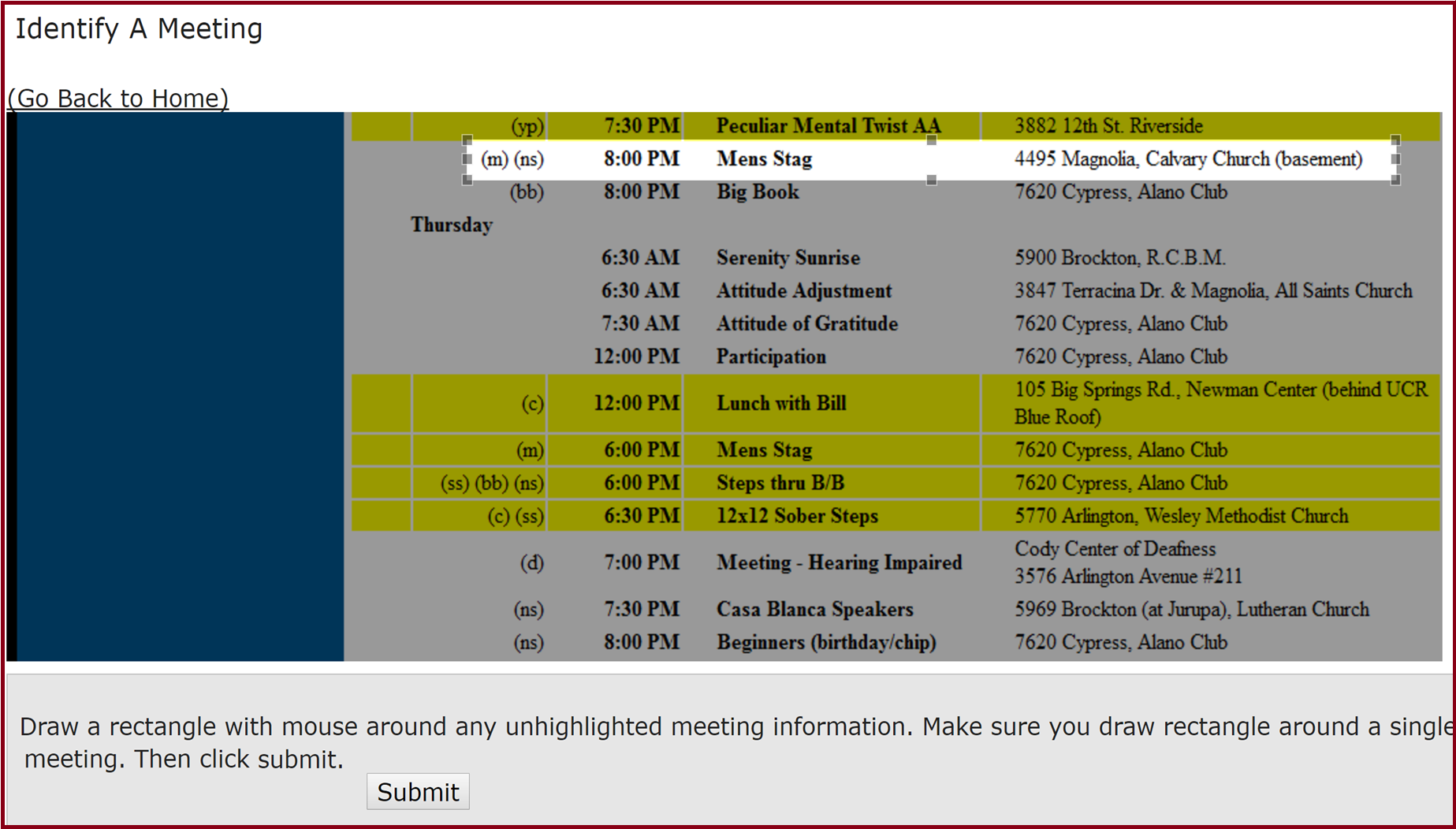}
    \caption*{(c)}
\end{figure}
\begin{figure}[h]
    \vspace*{-0.5cm}
   \caption{Interfaces of the three tasks (a) meeting page validation, (b) meeting validation, and (c) meeting identification}
 \label{fig:threetasks}
\end{figure}

\subsection{Tasks}
\label{sec:tasks}
In the context of recovery, the researchers in previous work turned to crowd workers when automated approaches failed to detect webpages containing meeting lists and to extract meeting patterns in them \cite{HAIR}. We compared the performances of the three different recruitment methods for the particular tasks of meeting page and meeting validation, which we divided into three different HITs for the workers.

We created responsive web interfaces for each task using the Python Flask framework. All workers were assigned to complete all three tasks, but we randomized the order of the tasks shown to them to reduce the learning effect on the results. From both MTurk and ITR platforms, they were redirected to the task website and were shown a consent form. Documentation of informed consent was waived in order to preserve participant anonymity. After they agreed to participate in the study, we showed them brief descriptions for each of the tasks. Prior to starting each task, there was a popup page with detailed instructions and examples (See supplementary document).

\subsubsection{Meeting Page Validation} The interface for page validation showed 10 webpages (selected randomly from the ground truth data set of pages) sequentially and asked the worker if it was a meeting page with \quotes{yes}, \quotes{no}, and \quotes{not sure} options. If workers selected \quotes{not sure,} they had to answer two or three additional questions to help us determine the label (Fig. \ref{fig:threetasks}a).

\subsubsection{Meeting Information Validation} The interface sequentially showed 10 meetings (selected randomly from the ground truth data set of meetings) highlighted in yellow on corresponding webpages and asked the worker if it was a meeting record, with \quotes{yes} and \quotes{no} options (Fig. \ref{fig:threetasks}b). If workers selected \quotes{no} for a meeting, they were advanced to the next record. Otherwise, they were prompted to edit the automatically extracted details of the meeting if these did not match with the highlighted record.

\subsubsection{Meeting Information Identification} The interface sequentially showed 10 meeting pages (selected randomly from the ground truth data set of pages) with the retrieved meetings highlighted in yellow and asked the workers if they noticed any meeting as not highlighted, with \quotes{yes} and \quotes{no} options. If workers selected \quotes{no} for a page, they were advanced to the next page. Otherwise, they would be asked to draw a rectangle around any one unhighlighted meeting (Fig. \ref{fig:threetasks}c).

Prior to recruitment, we conducted several pilot experiments: two on MTurk and two with undergraduate volunteers, to refine the interfaces of the tasks and task instructions and descriptions, to determine the appropriate pay per worker (\$2.5), and to define a reasonable minimum time of task completion (five minutes).

\subsection{Platform Cost}
Our research questions are related to calculating and comparing the cost of task completion for different worker types, that is the total cost paid to the workers only for completing the tasks, since this fraction of the total cost would more likely influence the workers' task completion time and accuracy. However, for this study, we had to pay additional costs for recruiting and filtering workers. In the interest of keeping the discussion about cost of the task completion consistent, we separate the platform cost that was associated with the use of particular crowdsourcing platforms to recruit workers, from the actual worker cost that was the amount of money that went to the workers who completed all the tasks. However, we acknowledge that in reality the required number of participants might not have been obtained without the advertising that caused the platform cost for recruiting the community-situated crowd. Therefore, we also discuss the total and average worker cost including all the expenses spent for the study (Section \ref{costtotal}).

On MTurk, the price a requester has to pay for a HIT is comprised of two components: the amount to pay workers plus a fee to pay MTurk. The usual MTurk fee is 20\% of the worker fee for any HIT. For each added worker qualification in this study (i.e., workers with greater than 90\% HIT acceptance and workers with correct answers on the screening survey) an additional 5\% of the worker fees was added to the platform cost.

For generic crowdsourcing on MTurk, the researchers paid a total platform cost of \$543.75 (25\% of the total worker cost described in the Results Section). This was a sum of the usual 20\% of the worker cost and the 5\% fee for an additional criterion of making the HIT visible only to workers having average HIT acceptance rate of 90\% or higher. For community-situated crowdsourcing on MTurk, the platform cost was a total of \$274.95, including the MTurk fees of \$4.95 for the screening survey and \$270 for the actual HIT. 

The additional cost of the workers filling out the screening survey could be reduced by including the survey as a part of the actual HITs, and then considering the HITs from only those workers with correct answers to the survey questions. However, that might not be a good idea for this research, assuming that plenty of workers who are not members of a 12-step fellowship would attempt the HITs but would get rejected due to low quality answers, increasing both worker dissatisfaction and our labor for data management and filtering. Considering the small amount of money and design effort introduced by the screening survey, we decided to keep it separate from the actual HIT.

We paid \$6000 to the founders of InTheRooms to run the advertisement for two months. In addition to putting up the banner on their homepage and on the website's weekly newsletters, the website founders guided us in improving the banner design and the task interfaces to attract more participants, and provided continuous support in recruiting participants through batch emailing community members about the research asking for their participation. For other domains, however, the platform cost would largely depend on the type of platform to recruit participants from and would vary based on the type of the tasks in question and the required community to perform them. We discuss this point in more detail in the discussion section.

\subsection{Measures}

For comparing the performance of different types of crowdsourcing, the variables we calculated and the statistical tests we conducted are as follows:

\begin{itemize}
    \item{Cost: We calculated unit and total costs paid to the workers and to the platforms.}
    
    \item{Time of recruitment: We report the amount of time required to recruit the required number of valid participants from each platform.}
    \item{Average accuracy and average task completion time: For the three different groups of workers, generic crowd, community-situated crowd from MTurk, and community-situated crowd from ITR, we at first conducted Kruskal-Wallis tests for mean accuracy and mean completion time for page validation, meeting validation, and meeting identification tasks to find out if there was a difference between at least one group and the other groups of workers. Afterwards, we performed between-subjects Wilcoxon Rank-Sum Tests (as some of those distributions were not normal) for only the tasks that had $p$-values less than 0.05 from the Kruskal-Wallis tests to calculate the difference in mean accuracy and mean completion time for the three pairs: generic crowd and community-situated crowd from MTurk, generic crowd from MTurk and community-situated crowd from ITR, and community situated crowd from MTurk and ITR. To account for multiple comparisons, we adjusted the $p$-values using the Holm-Bonferroni method.}
    Since the type of data to identify and validate was different in each task, we calculated the accuracy differently.
    $$\text{Accuracy}_{page\_validation} = \frac{ \text{\# of pages correctly answered with yes/no}}{\text{total \# of pages validated with yes/no}}$$
    
    $$\text{Accuracy}_{meeting\_validation} = \frac{ \text{\# of meetings correctly validated (labeled and edited)}}{\text{total \# of meetings attempted}}$$
    
    $$\text{Accuracy}_{meeting\_identification} = \frac{\text{\# of meetings correctly identified (labeled and highlighted)}}{\text{total \# of meeting pages attempted}}$$
    
    \item{To compare the performances for specific portions of the tasks that require workers' contextual knowledge, we conducted appropriate statistical analyses. For instance, to calculate the difference in average number of meetings validated with a \quotes{not sure} option, we conducted a Chi-Squared test, where the categories were represented as whether a particular page was classified with a \quotes{not sure} option or not by a particular type of crowd worker. }
    
\end{itemize}

\section{Results}
We analyzed the performance of generic crowdsourcing and community-situated crowdsourcing with workers from two different platforms and discuss the trade-offs in this section in terms of recruitment time and cost, task completion time, and accuracy of the workers. Additionally, we were interested in understanding if community-situated crowd workers performed better in detecting and validating particular types of information due to their familiarity with the context.

\begin{table}[]
\centering
\resizebox{\textwidth}{!}{%
\begin{tabular}{@{}lcc@{}}
\toprule
 &
  \begin{tabular}[c]{@{}c@{}}Worker cost only \\ (average cost per worker)\end{tabular} &
  \begin{tabular}[c]{@{}c@{}}Total worker cost including platform cost\\ (average cost per worker)\end{tabular} \\ \midrule
Generic: MTurk            & \$2175 (\$2.5)                              & \$2718.75 (\$3.125) \\
Community-situated: MTurk & \$1125 (\$2.5) + \$19.80 (\$0.03) = \$1099.80 & \$1431 (\$3.18)     \\
Community-situated: ITR   & \$0 (\$0)                                   & \$6000 (\$13.33)    \\ \bottomrule
\end{tabular}%
}
\caption{Worker costs and total costs for generic and community-situated crowdsourcing. Worker cost is the portion of the total cost paid to only the workers for their task completion, and not to the platform.}
\label{tab:workerandplatformcost}
\end{table}

\subsection{Cost of the Workers}\label{costtotal}
As mentioned earlier, worker cost is the portion of the total cost that went to the workers, and not to the platform. For the MTurk HIT to complete the tasks, \$2.50 was paid to each worker. For generic crowdsourcing in the previous work, 870 HITs were accepted and we accepted 426  HITs from the community-situated crowd answers. Since the total number of participants in these two samples are different and there was an additional cost of \$19.80 for the 660 accepted HITs of the screening survey in the MTurk community-situated crowdsourcing, there is a difference in the total worker cost between generic workers and community-situated MTurk workers. The total worker cost is \$2175 (\$2.50 per worker) and \$1099.80 (~\$2.57 per worker) for generic MTurk and community-situated MTurk workers respectively. We did not pay any monetary rewards to the ITR workers. Therefore, considering the \textit{average cost per worker}, community volunteers cost the least (zero), followed by generic crowdsourcing from MTurk, and community-situated crowdsourcing from MTurk respectively.
Table \ref{tab:workerandplatformcost} reports these worker costs.

Ideally, if the same number of paid generic and community-situated crowd workers are recruited to perform the same number of tasks, the cost would be about the same. In contrast to the paid community-situated MTurkers, the community-situated worker cost can potentially be very low or even zero, irrespective of the number of recruited  workers, if we choose to recruit volunteers from online groups/communities relevant to the context in question. However, these calculations disregard the platform costs which can significantly increase the average cost per worker for community-situated workers in many studies similar to this one. We discuss the total cost and the average worker cost including all platform costs in the next subsection.



\begin{table}[]
\centering
\begin{tabular}{@{}lcc@{}}
\toprule
{\color[HTML]{000000} }                       & {\color[HTML]{000000} Time}                                              & {\color[HTML]{000000} Accuracy} \\ \midrule
{\color[HTML]{000000} Page validation} &
  {\color[HTML]{000000} \begin{tabular}[c]{@{}c@{}}\textless{}.001\\ ***\end{tabular}} &
  {\color[HTML]{000000} \begin{tabular}[c]{@{}c@{}}\textless{}.001\\ ***\end{tabular}} \\
{\color[HTML]{000000} Meeting validation} &
  {\color[HTML]{000000} \begin{tabular}[c]{@{}c@{}}\textless{}.001\\ ***\end{tabular}} &
  {\color[HTML]{000000} \begin{tabular}[c]{@{}c@{}}\textless{}.001\\ ***\end{tabular}} \\
{\color[HTML]{000000} Meeting identification} & {\color[HTML]{000000} \begin{tabular}[c]{@{}c@{}}.003\\ **\end{tabular}} & {\color[HTML]{000000} .172}     \\ \bottomrule
\end{tabular}
\caption{$p$-values from the Kruskal-Wallis tests for the differences in task completion time and accuracy among the three tasks}
\label{tab:kwtimeandaccuracy}
\end{table}

\subsubsection{Total Cost}
\label{sec:totalcost}
Even though in our study the platform cost ideally did not impact  the workers' task completion time and accuracy, many studies may require high advertising costs to ensure enough participation and/or to obtain quality results. The third column of Table \ref{tab:workerandplatformcost} represents the total and the average costs including both worker and platform costs. The average cost per worker for generic and community-situated crowdsourcing were \$3.125 and \$3.18 respectively. The additional average worker cost for community-situated MTurk is due to the HIT used for filtering workers. The average worker cost for ITR participants was \$13.33 considering the platform cost, which is very expensive compared to the other two populations. Including platform cost in fact reorders the worker cost by type required for this study. However, for tasks related to other contexts/domains, or even in the context of AA, the platform cost can be reduced by recruiting workers from other platforms with large user base. We discuss it in more details in Section \ref{sec:impresearch}.

\subsection{Time to Recruit Workers}
MTurk and ITR both have over 500,000 members, although not all of them are active \cite{pewdemomturk}. Paid online platforms like MTurk have been shown to be an effective way of quickly recruiting workers (both experts and novices) \cite{borromeo2016investigation}. We found evidence for this in our study, as it took the least time to recruit generic workers from MTurk.

For generic crowdsourcing, the researchers aimed to recruit participants until they achieved validation results for a predetermined number of meeting pages and meetings \cite{HAIR}. They published HITs in batches and received the desired number of responses in about a week ($n=1060$). For community-situated MTurkers, it took two weeks to recruit 450 participants, starting from the time the screening survey HIT was published. For the ITR advertisement, we achieved a total of $460$ complete responses which was about the same number as the targeted number of participants for this comparison. It took, however, additional time for us prior to recruitment to negotiate with the website founders about the logistics of the advertisement. Therefore, time to recruit generic crowd workers was substantially less than both types of community-situated workers. Additionally, recruitment time of unpaid volunteers from ITR was about four times slower than paid community-situated workers on MTurk.  

In general for other domains, recruiting volunteers from online communities might consist of similar steps and might take substantially more time than recruiting paid community-situated workers. On the contrary, depending on the type of data validation, other platforms with more volunteers (e.g., Facebook groups) may be leveraged to reduce this time. Therefore, the recruiting organizations have to consider the trade-off between time and number of participants for different platforms based on the emergency and frequency of the crowdsourcing task in question (more about this in the discussion section).

\begin{table}[]
\centering

\begin{tabular}{@{}lcccccc@{}}
\toprule
\multirow{2}{*}{} & \multicolumn{2}{l}{Page validation} & \multicolumn{2}{l}{Meeting validation} & \multicolumn{2}{l}{Meeting identification} \\ \cmidrule(l){2-7} 
                 & M      & SD     & M      & SD     & M      & SD     \\ \midrule
Generic: MTtuk   & 237.61 & 162.11 & 318.37 & 255.50 & 268.50 & 212.78 \\
Community: MTurk & 315.12 & 181.70 & 555.73 & 332.46 & 309.49 & 313.62 \\
Community: ITR   & 343.23 & 265.59 & 507.27 & 360.37 & 329.26 & 316.08 \\ \bottomrule
\end{tabular}%

\caption{Means and standard deviations of task completion time (in seconds) of generic and community-situated crowd workers for three different tasks}
\label{tab:workertimecompletion}
\end{table}

\begin{table}[]
\centering
\resizebox{\textwidth}{!}{%
\begin{tabular}{@{}lcccccc@{}}
\toprule
\multirow{2}{*}{} &
  \multicolumn{2}{c}{Page validation} &
  \multicolumn{2}{c}{Meeting validation} &
  \multicolumn{2}{c}{Meeting identification} \\ \cmidrule(l){2-7} 
 &
  Time &
  Accuracy &
  Time &
  Accuracy &
  Time &
  Accuracy \\ \midrule
Generic and community: MTtuk &
  \begin{tabular}[c]{@{}c@{}}\textless{}.001\\ ***\end{tabular} &
  \begin{tabular}[c]{@{}c@{}}.021\\ *\end{tabular} &
  \begin{tabular}[c]{@{}c@{}}\textless{}.001\\ ***\end{tabular} &
  \begin{tabular}[c]{@{}c@{}}.002\\ **\end{tabular} &
  .18 &
  .15 \\
Generic and community: ITR &
  \begin{tabular}[c]{@{}c@{}}\textless{}.001\\ ***\end{tabular} &
  \begin{tabular}[c]{@{}c@{}}\textless{}.001\\ ***\end{tabular} &
  \begin{tabular}[c]{@{}c@{}}\textless{}.001\\ ***\end{tabular} &
  \begin{tabular}[c]{@{}c@{}}\textless{}.001\\ ***\end{tabular} &
  \begin{tabular}[c]{@{}c@{}}.030\\ *\end{tabular} &
  .15 \\
Community: MTurk and community: ITR &
  .97 &
  \begin{tabular}[c]{@{}c@{}}\textless{}.001\\ ***\end{tabular} &
  .163 &
  .051 &
  .25 &
  .87 \\ \bottomrule
\end{tabular}%
}
\caption{Statistical comparisons of average task completion time and accuracy between: 1) Generic crowdsourcing and community-situated crowdsourcing from MTurk, 2) Generic crowdsourcing and community-situated crowdsourcing from MTurk, and 3) Community-situated crowdsourcing from MTurk and ITR. $p$-values are reported using Wilcoxon Rank-Sum Tests and adjustment with Holm-Bonferroni method. (* $p<0.05$, ** $p<0.01$, *** $p<0.001$)}
\label{tab:timeandaccuracycomparison}
\end{table}

\subsection{Task Completion Accuracy of Workers}
\begin{center}
\begin{figure}[t]
\begin{tabular}{ l l l }
\begin{subfigure}{0.33\textwidth}
      \includegraphics[width=\textwidth]{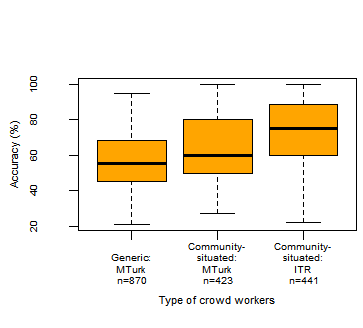}
      \caption{}
      \label{box1}
    \end{subfigure}    &
    
    \begin{subfigure}{0.33\textwidth}
      \includegraphics[width=\textwidth]{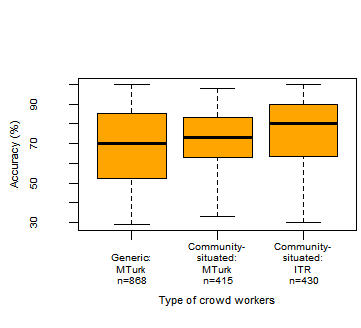}
      \caption{}
      \label{box2}
    \end{subfigure} 
    
    \begin{subfigure}{0.33\textwidth}
      \includegraphics[width=\textwidth]{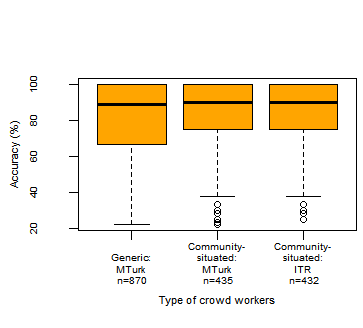}
      \caption{}
      \label{box3}
    \end{subfigure}
\end{tabular}
 \caption{Boxplots of accuracy achieved by three different types of crowd workers for: (a) page validation, (b) meeting validation, and (c) meeting identification}
 \label{fig:accuracyfortasks}
\end{figure}
\end{center}

InTheRooms volunteers completed the tasks with highest accuracy regardless of the type of the task, followed by community-situated MTurkers and generic MTurkers, respectively (refer to the boxplots in Fig. \ref{fig:accuracyfortasks}). Kruskal-Wallis tests showed significant difference in mean accuracy except for the task of meeting identification (Table \ref{tab:kwtimeandaccuracy}). For page validation and meeting validation, the differences between generic crowd workers and the community-situated crowd workers were statistically significant at the 95\% confidence level. Additionally, mean accuracy of the community-situated workers from ITR was significantly higher for page validation than community-situated workers from MTurk. For meeting validation, accuracy of both types of community-situated workers were significantly higher than generic crowd workers. We discuss these accuracies and their interpretations in more details in the following subsections.
\subsubsection{Page Validation}
The mean accuracy of community-situated crowdsourcing from ITR ($M=72.1\%,\ SD=19.18\%$) was significantly higher than both the mean accuracy of community-situated crowdsourcing from MTurk ($M=60.52\%,\ SD= 18.95\%$) ($p<.0001$) and generic crowdsourcing from MTurk ($M=56.40\%,\ SD=16.24\%$)($p<.0001$) (Fig. \ref{box1}). This difference in accuracy in the context of creating an AA meeting list means that recruiting participants from online communities for recovering alcoholics can increase the accuracy of meeting page validation by 16\% compared to a generic platform like MTurk, resulting in correct label for about one in six additional meeting pages in the data set.

\subsubsection{Meeting Validation}
Similar to the task of page validation, the highest mean accuracy ($M=76.94\%,\ SD=17.33\%$) was achieved by the community-situated workers from ITR, followed by community-situated MTurkers ($M=73.54\%,\ SD=15.45\%$) and generic MTurkers ($M=71.54\%,\ SD=20.51\%$). This accuracy was significantly higher than generic crowdsourcing from MTurk ($p<.0001$) (Fig. \ref{box2}). In addition, community-situated MTurkers' accuracy was significantly higher ($p$=.002) than the accuracy of generic MTurkers. In the context of creating an AA meeting list, this finding implies that the ITR members can provide accurate day, time, and address information for about 5\% more meetings than the generic crowd workers. AA hosts more than 50,000 meetings throughout the United States and community-situated volunteers can substantially improve the reliability of the meeting list by validating these meetings accurately.

\subsubsection{Meeting Identification}
Community situated ITR workers achieved the highest mean accuracy of meeting identification task ($M=81.80\%,\ SD=21.24\%$), followed by community-situated MTurk workers ($M=81.36\%,\ SD=25.39\%$) and generic MTurk workers ($M=79.76\%,\ SD=23.36\%$) respectively (Fig. \ref{box3}). There was no significant difference in accuracy between any of the three pairs. We assume that, since meeting identification involved only segmenting a part of a webpage if there was a meeting that was not highlighted (i.e., drawing a rectangle around an unhighlighted meeting record), it is comparable to the task of segmenting images for object detection and is a common type of HIT on crowdsourcing platforms. In fact, generic crowd workers completed this task with comparable accuracy in less time on average.

\begin{center}
\begin{figure}[t]
     \includegraphics[width=0.7\textwidth]{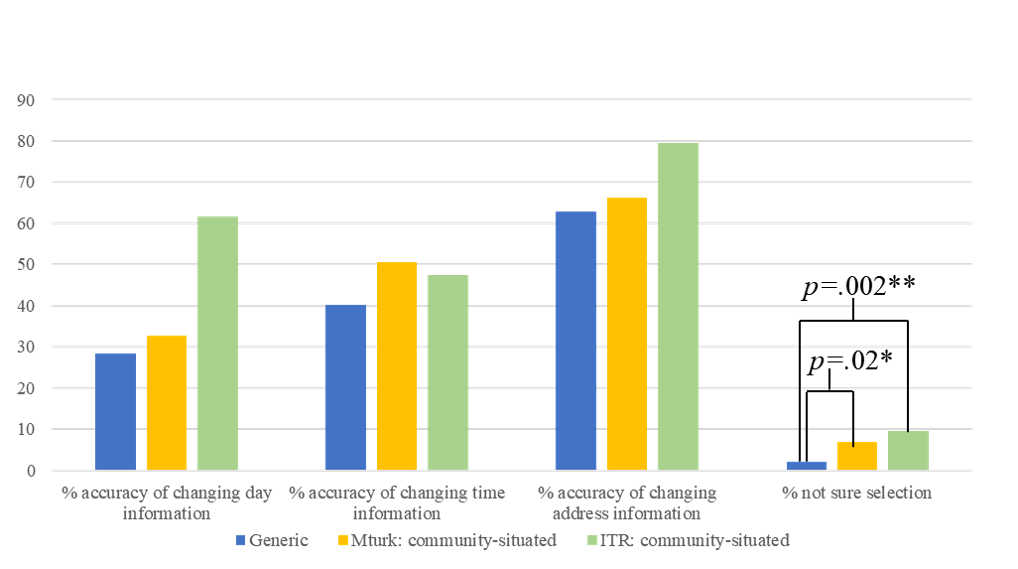}
\caption{Different types of errors in the page validation and the meeting validation tasks; the first three groups from the left respectively show the average accuracy of editing the day, time, and address information correctly in the meeting validation task. The rightmost group shows the the average percentage (and the differences between groups) of selecting the \quotes{not sure} option in the page validation task}
\label{tab:barchart}
\end{figure}
\end{center}
\subsection{Evidence for Reasons of Difference in Accuracy}
We were interested to further investigate the possible reasons for the significant difference in accuracy between community-situated crowdsourcing and generic crowdsourcing. Through post-hoc analyses and a closer look into the data, we found out that this difference in accuracy could be contributed by one or more of the following factors:

\subsubsection{Task Completion Time of Workers}
The average time taken to complete the tasks by generic MTurkers (page validation: $M=237.61s,\ SD=162.11s$, meeting validation: $M=318.37s,\ SD=255.50s$, meeting identification: $M=268.50s,\ SD=212.78s$) was less than both the community-situated MTurkers (page validation: $M=315.12s,\ SD=181.70s$, meeting validation: $M=555.73s,\ SD=332.46s$, meeting identification: $M=309.49s,\ SD=313.62s$), and the ITR workers (page validation: $M=343.23s,\ SD=265.59s$, meeting validation: $M=555.73s,\ SD=332.46s$, meeting identification: $M=329.26s,\ SD=316.08s$). Generic crowd workers completed the tasks of page validation and meeting validation significantly faster than both the paid ($p<.001$) and unpaid ($p<.001$) community-situated crowds. For meeting identification, this difference was not significant. There was no significant difference in completion time of any of the three tasks between paid and unpaid community-situated worker samples (see Time columns in Table \ref{tab:timeandaccuracycomparison}).

In terms of average completion time by type of task, page validation took the least amount of time for all types of workers and meeting validation took the most. This is expected, since for validating a meeting workers often have to scroll through the webpage shown to them to confirm that the time and location of the highlighted meeting is accurate, and edit the information if necessary. On the other hand, for validating a meeting page they mostly need to skim through the page shown to them for a list of meetings and determine if those are open AA meetings. Unlike editing and typing texts in meeting validation task, for page validation they had to answer one (a maximum of 2-3 if they selected \quotes{not sure}) multiple choice question.

These results indicate that workers being paid for task completion were probably performing the tasks very quickly and paying minimal attention to each one, causing the lowest accuracy for generic MTurkers. This finding also resonates with results from many previous studies on paid crowdsourcing \cite{mao2013a, borromeo2016investigation}. For the community-situated crowd samples, volunteers from ITR usually took slightly more time on average than the paid MTurkers. This could probably be attributed to the different motivations of the community-situated crowd from these platforms: For ITR members the only reason of participation was their personal motivations to be of service to the community, whereas MTurk members were both members of 12-step fellowships (supposedly) who wanted to help other community members, and also belonged to online crowd workers seeking to maximize their earnings through completing HITs quickly.

\subsubsection{Knowing When \quotes{Not Sure}}
We calculated the reported accuracy of page validation considering only those workers who were certain in labeling most of the webpages shown to them. To be more specific, if a worker selected \quotes{not sure} for more than three pages, we did not include his/her response to calculate the accuracy for this particular task. We assumed that the accuracy difference may occur partly because of the willingness of different types of workers in choosing the \quotes{not sure} option when they were actually not sure. To find out if this assumption was true, we calculated the average number of pages where workers selected the \quotes{not sure} option, and this was highest for the ITR members ($M=9.48\%$) and lowest for the generic MTurkers ($M=2.1\%$) (Fig. \ref{tab:barchart}). This measure was calculated by counting the number of pages where not sure option was selected by a particular type of workers and dividing that number by the total number of pages shown to all workers of that type. Since ITR workers achieved the highest average accuracy for page detection, we assume that the community members are more confident of what they do not know and they did not want to provide incorrect answers (more on Section \ref{sec:impresearch}). A Chi-Square test showed that this number was significantly higher for ITR than generic MTurkers (${p}=.002$). 

In the cases where workers were not sure, they could choose to answer a 2-3 additional questions that could guide us in determining the probable label of the webpage. Through an analysis of the answers to those additional questions we identified that ITR workers were more willing and accurate in answering those additional questions. In other words, to calculate average accuracy, if we consider the pages where we can obtain a label through workers' answers to additional questions, we get a 1\% increase in the accuracy for ITR workers, whereas the accuracy of generic and community-situated MTurkers do not change at all.

\subsubsection{Better Context Interpretations}
We suspected that the community members could perform better in differentiating meeting pages from other pages due to their background knowledge of the AA program's structures and norms. For example, the findings from the previous work with generic crowdsourcing suggested for page validation that, many generic workers labeled webpages with events or office hours (that have times and locations) inaccurately as meeting pages \cite{HAIR}. We assumed that the community-situated workers would label pages more accurately in these cases. However, we did not have explicit ground truth data for a set of pages that listed events or office hours, and therefore, we cannot claim that ITR workers were significantly more accurate in differentiating meeting and event pages than other types for workers. However, we manually checked some of the pages that were labeled differently by generic and community-situated workers and found that some event pages were correctly classified as non-meeting pages by community-situated  workers only, possibly contributing to a higher accuracy of the ITR workers. Similarly, based on our observation, community-situated workers performed better than generic crowd workers in differentiating single event information (e.g., day, time, address of picnics or service meetings, office hours etc.) from meeting information.

For the task of meeting validation, total accuracy of a worker was attributed to being able to identify whether the highlighted information is a single meeting (i.e., it includes only one day, one time, and one address of an AA meeting)  and to edit the day, time, and address accurately to match the ground truth. Due to better context interpretations of the community-situated workers, we assumed that there may be difference in how different types of workers edit different information.  Fig. \ref{tab:barchart} shows the average percentage of day, time, and address information edited by the workers. The higher accuracy of the community workers than the generic workers in editing the day information might occur due to workers being more familiar with the structure of the webpages and taking more time to skim through the pages. This is because from our observations, we had seen many meeting websites that listed the meetings per day of the week. For these types of pages, workers had to scroll through the top of the page to edit the day information.

\section{Discussion}
The implications of these results point to next steps in our research on providing a reliable information source for meetings for people in recovery. In addition, we discuss practical implications for technology design given the impacts of worker type on the outcome of data validation, as well as research opportunities for the CHI community.
\subsection{Implications for the Context of Recovery}
\label{sec:imprecovery}
Our investigations point to the importance of considering the trade-offs of time, cost, and accuracy while adapting crowdsourcing for AA meeting data validation. Finding a meeting is currently a challenge for many AA members. The list of meetings in an area largely depends on local AA groups continually updating changes to meetings, which makes the current meeting finders outdated for different regions. Crowdsourcing, when combined with automated information retrieval techniques, can help periodically extract and validate meeting information from different websites and provide a reliable up-to-date source of information. We used the context of recovery as an example to broadly answer our research questions regarding community-situated crowdsourcing applicable to other domains as well. Consequently, our results had many interesting implications for the next step in utilizing crowdsourcing effectively in this specific domain.

InTheRooms volunteers achieved the highest accuracy for most of the tasks (including the comparatively longer task of meeting validation) among the three different worker samples. This finding implies that we can rely on volunteers for data validation for this particular problem. This resonates with the findings of many other studies showing that community-based crowdsourcing results in more accurate and reliable outcomes \cite{community1, communitycontribute, citizenscience}. Keeping the meeting lists up-to-date would require help from workers in periodic intervals and community-situated crowdsourcing can ensure a stream of continuous workers, potentially reducing the worker cost to zero or close to that.
This particular study caused us a higher platform cost for recruiting ITR members than both generic and community-situated MTurkers. Although the platform cost does not impact worker performance directly, this is part of the total cost required for the study. This platform cost can be significantly reduced both by using other online platforms and apps, and by developing our own system to recruit community members. For example, with the meeting data initially extracted and validated, we can build an up-to-date meeting finder application targeted for AA members. When the app is in use, we can potentially ask the users of the app to volunteer in helping us improve the meeting list through validating meeting information. However, relying on a system or app in its initial phase with a small number of users may be questionable to recruit volunteers, since depending on the volume of data requiring validation, recruiting the desired number of participants can take a substantial amount of time. For example, we recruited from an online community with more than 500,000 members, but the recruitment time was about two months for 450 participants. As a counter-argument, our study limited the participants to perform the tasks only once in order to remove the bias of learning by repetition, whereas in a real-time system, each worker would be able to perform the tasks as many times as possible, minimizing the trade-off in recruitment time, and improving worker expertise at the same time.

In summary, volunteers from recovery communities can be relied upon for validating AA meeting webpages and meetings, and the cost of recruitment can be substantially reduced by careful consideration in adapting different approaches to attract more volunteers.

\subsection{Implications for Design of Crowdsourcing ICT Systems}
Our findings suggest that many design decisions of crowdsourcing ICT systems, such as the task interfaces, the choice of platform, etc. should consider the
trade-offs in cost and accuracy. We found statistically significant difference in the mean percentage of accuracy between the generic and the community-situated crowd workers. However, \quotes{statistically significant} differences in time, accuracy, or cost between different types of crowdsourcing may not be practically significant for many other domains. Careful considerations have to be made regarding these trade-offs to select the type of workers and the platform depending on the monetary budget, urgency of retrieving crowdsourced answers, and the accepted threshold of errors in those.

For example, the choice of a platform to solicit the community-situated workers from largely depends on the budget for the platform cost, existing collaboration with the platform's owners, and the sensitivity of the context. In the case of recovery, members are particularly vulnerable and anonymity is of utmost importance to them \cite{rubyaanonymity}. We made sure to design the tasks in a way such that no identifiable personal information has to be revealed by the participants. If the information being validated requires crowdsourcing from people with a stigmatized health condition, members of online communities where real identity is associated (e.g., Facebook health support groups) may not show enough interest in participation. Additionally, if the data requires validation in frequent interval like the context of this study, requesters have to choose a platform from where they can recruit the expected number of workers within the time constraints. While designing tasks in this type of recruitment, one or more existing techniques of minimizing errors while getting rapid answers \cite{kittur2008crowdsourcing, nath_threats_2012, embracingerror} should be adopted.

About 4\% of our participants from ITR completed the tasks partially (e.g., completed one or two tasks). The order of the tasks was randomized but a close look at our data revealed that many of them left in the middle when the task of meeting validation was shown to them as the last task. From our results, we found out that validating meetings took the longest time on an average among the three tasks for both types of crowdsourcing. Making the overall task more granular (i.e., one task per worker instead of three) would probably yield more workers completing the assigned task and increase engagement, as suggested by previous work \cite{dynamicsmicrotask,turkomatic}. However, in the case of ITR workers, who were redirected to the task interfaces from clicking on the advertisement, dividing the tasks would require three times the number of interested participants (about 450 participants per task). While designing task interfaces, these trade-offs are important to consider.
\subsection{Implications for Research}
\label{sec:impresearch}
Findings from our study provide practical implications for further research regarding the trade-offs and benefits of generic vs. paid community-situated vs. unpaid community-situated crowdsourcing in different contexts. One of the important factors influencing this choice is the type of task to be crowdsourced. For example, generic MTurkers in this study achieved about the same accuracy for meeting identification, but were significantly quicker in completing the task. However, for the other two tasks, their accuracy was way lower than the community-situated crowd. Therefore, piloting the tasks with different types of workers prior to running the actual experiments may provide a guideline for determining which tasks are more suitable for a particular type of crowdsourcing vs. the others. Similarly, the choice of recruiting community-situated workers from online marketplaces like MTurk rather than online communities can be influenced by the actual proportion of the community members in the general population. For instance, about 1\% Americans are members of AA \cite{aamembership}, and MTurk typically represents similar proportions of particular community members as the general US population. In many other domains, however, the required community members may represent a larger proportion of the general us population (e.g., students). Researchers conducting experiments related to those communities may seek community-situated crowd workers from MTurk or other online marketplaces, where they can recruit a large number of people very quickly, instead of selecting online groups with comparatively smaller number of people (e.g., online communities of students or public Facebook groups, etc.).

The choice of platform has an impact on the platform cost as well. Our study incurred a high platform cost for advertising on ITR. As discussed earlier in this paper (Section \ref{sec:imprecovery}), the survey can be integrated in external apps and websites related to recovery reducing the total cost per worker. Similarly, for other studies researchers can optimize the platform cost by carefully choosing a platform that has no or little impact on other performance measures (e.g., time of recruitment, answer quality, etc.). Let's take Facebook advertising as an example. The average cost per click for Facebook ads is \$1.72 \cite{facebook_nodate}, and thus using Facebook ads instead of ITR would possibly reduce the platform cost for this study. However, it might have impacted the response rate, as AA members are particularly concerned about their anonymity \cite{rubyaanonymity}, and there may not be a large number of targeted 12-step members on Facebook who would participate. This may not be an issue for studies requiring other types of community-situated workers, and those may benefit from recruiting workers through platforms with lower advertising costs. Future research should aim to understand the relationship between platform cost and worker performance, and provide a comprehensive guideline on selecting the right platform for a study.

Research focused on online community-based crowdsourcing should address particular motivations and expertise of the community members to figure out the feasibility and effectiveness of recruiting such workers. HCI researchers have pointed out that community-based intrinsic motivations such as altruism, collectivism, reciprocity, etc. play an important role in worker participation and engagement \cite{collectivecreativity, socialpurpose, crisisdisaster, ideasharing, alam_crowdsourcing_2012, Kaufmann2011More}. Similarly, worker expertise and knowledge about the problem domain have proven to be directly related with the quality of crowdsourced data \cite{almostexpert, Vries2017Experts, warby2014a}. In many other contexts like recovery, the expertise can come from particular community members who are willing to volunteer for social reasons, such as to achieve recognition in the community or to provide service. Researchers should investigate generic and community-situated crowdsourcing in those domains to understand the benefits and the trade-offs in time, quality, cost, etc. For example, members of an online community of cancer caregivers may provide better answers to a care-giving related question than a non-caregiver.

While community members are often considered as experts for the target task, recruiting such participants from paid online platforms has the drawback of workers falsely reporting their membership. Although screening questions can reduce the effect, there is always a possibility that some workers have made their way to bypass the screening techniques. Further research can focus on analyzing the impact of different screening techniques on the worker performance while recruiting from paid platforms.

ITR members in our study did not receive any monetary incentives. We assume a major motivation behind their participation was to provide service to the community by helping us create a useful resource for the newcomers in the community. We provided an option for the workers to leave us comments on the task descriptions and the design, and some of the comments from the ITR participants actually revealed this \quotes{sense of providing service} to the community by helping others (e.g., \quotes{thank you for allowing me to help out.I believe it's our purpose to help the newcomer if anyone needs help.}). This intrinsic motivation probably also encouraged them to take sufficient time to select accurate answers for as many questions as they could. One might expect the community-situated MTurkers to perform better than ITR members, as they should have the same intrinsic motivation along with monetary remuneration for task completion. However, they might not have paid enough attention in order to complete more tasks in less time, resulting in lower accuracy of task completion than the ITR participants. Prior research has established the impact of incentives and motivations on not only who participates in a particular task \cite{youget}, but also on their attention, dropouts, and performance \cite{paybackward, motivation1}. Based on prior work and our findings that different types of motivations and incentives may affect workers' attention and question-answering behavior differently, we recommend future research to use motivation and incentives to analyze reliability of the answers. As an example, whenever possible, researchers can compute worker motivation using established scales \cite{measuringmotivation2019} and figure out if workers with particular motivation types provide quality answers and further target workers with similar motivation.

Using the wisdom and motivations from community-situated crowds in this context produced high quality results. This provides implications to develop approaches for collaborative crowdsourcing, pairing generic workers with community-situated workers. In the context of paid microtasks, collective crowdsourcing with the paired-worker model has led to better accuracy and increased output, which, in turn, has translated into lower costs \cite{socialcollaborative, leasework}. Moreover, different types of workers play different roles and exhibit different potentials in crowdsourcing \cite{websecurity, Huang2017Leveraging}. Researchers can build on the findings of these previous works and our study to create coordination, where community-situated workers help set directions and guide the non-expert generic workers.

The number of ITR members selecting the \quotes{not sure} option for the page validation task was significantly higher than the other two populations. This finding echoes the well known Dunning–Kruger effect, which is defined in psychology as a cognitive bias in which low-ability individuals suffer from illusory superiority, mistakenly assessing their ability as much higher than it really is \cite{dunning_chapter_2011}. In other words, experts tend to know what they do not know. In the context of crowdsourcing, this may affect answer quality tremendously if the task comprises many difficult or confusing questions. Therefore, further research needs to be done to find out ways to minimize this effect. Overall, our findings and collective observations highlight multiple opportunities and directions with pursuing deeper understanding of effective applications of community-situated crowdsourcing in different domains, not compromising the quality of the crowdsourced results.
\subsection{Broader Impact and Takeaways}
Our study results have practical implications regarding performance trade-offs for different types of crowdsourcing that may be helpful for other HCI and CSCW researchers. It discusses important factors to consider in determining platforms, worker types, and incentives for crowdsourcing. The results suggest that generic crowdsourcing platforms are good for some tasks on a specific domain while other seemingly similar tasks will benefit from domain knowledge. Perhaps this can inform the creation of more complicated setups, where an initial screening can help direct the workers to the right kind of task, even in the context of the same task. Future research should also benefit from the implications regarding the two types of crowdsourcing working in tandem.
\section{Limitations}
We assumed that the community members from InTheRooms are in fact members of 12-step fellowships and did not provide the screening questions for them. Although seemingly there is no particular reason for a non-member to join this online community and take part in the study (as discussed in Section \ref{sec:itrworkers}), using the same screening criterion for both platforms would make the study conditions fairer. Similarly, even though the three questions we used for screening were not too easy to bypass and we adopted additional filtering criteria to approve answers, we cannot certainly claim that all of the community-situated MTurkers were in fact 12-step fellowship members. In addition, the motivation and the incentive (i.e., pay for the MTurk community-situated workers, altruism and sense of belonging to the community for ITR workers, etc.) may have impacted the outcome of the study, and therefore, follow-up experiments should be conducted to rule out the impact of incentive and motivation. In particular, comparisons of task completion time, accuracy, cost, etc. should be carried out with sample workers recruited from ITR with monetary incentive and generic workers recruited for free.

\section{Conclusion} 
In the context of crowdsourcing data validation for Alcoholics Anonymous meetings, we provided an empirical comparison of accuracy, cost, and time trade-offs in generic crowdsourcing from MTurk, community-situated crowdsourcing from MTurk, and community-situated crowdsourcing from an online recovery community, InTheRooms. Our results show that community-situated workers from the online recovery community achieved significantly higher accuracy in the validation tasks than the other two types of crowd workers. We further investigated the possible reasons for this difference in accuracy and found that task completion time, knowing when \quotes{not sure}, better context interpretations, etc. may have influenced the accuracy of different types of crowd workers. From our findings, we provide practical implications for crowdsourcing in the recovery context, as well as for research and design in other domains. We discuss the factors that should be considered while selecting community-situated workers from a particular platform vs. the others, including platform cost, urgency and frequency of crowdsourcing, type of the crowdsourced tasks, etc.
\begin{acks}
Thank you to the participants for their time and contribution. We thank Zachary Levonian
and Daniel Kluver for their useful feedback on the quantitative analysis. This work was funded by the NSF grants (1464376 and 1651575). 
\end{acks}

\bibliographystyle{ACM-Reference-Format}
\bibliography{references,manual-entries}


\end{document}